\documentclass[twocolumn,secnumarabic,amssymb,amsmath,
nofootinbib,tightenlines,showpacs,floatfix,
superscriptaddress,aps,prl]{revtex4-1}
\usepackage{amsfonts}
\usepackage{txfonts} 
\usepackage{amsbsy}
\usepackage{bm}%
\usepackage[colorlinks=true,linkcolor=blue,filecolor=blue,menucolor=yellow,urlcolor=blue,citecolor=blue,anchorcolor=blue]{hyperref}
\usepackage{graphicx}
\usepackage{dcolumn}
\usepackage{times} 

\begin{document}

\title{Reentrant superconductivity in conical-ferromagnet/superconductor nanostructures}  

\author{Chien-Te Wu}
\email{wu@physics.umn.edu}
\author{Oriol T. Valls}
\email{otvalls@umn.edu}
\altaffiliation{Also at Minnesota Supercomputer Institute, University of Minnesota,
Minneapolis, Minnesota 55455}
\affiliation{School of Physics and Astronomy, University of Minnesota, 
Minneapolis, Minnesota 55455}

\author{Klaus Halterman }
\email{klaus.halterman@navy.mil}
\affiliation{Michelson Lab, Physics
Division, Naval Air Warfare Center, China Lake, California 93555}

\date{\today}

\begin{abstract} 
We study a bilayer consisting of an ordinary superconductor and
a  magnet with a spiral magnetic
structure of the {\rm Ho} type. We use a self consistent
solution of the Bogolioubov-de Gennes equations to evaluate the 
pair amplitude, the transition temperature, and the thermodynamic
functions, namely, the free energy and entropy.
We find that for a range of thicknesses of the magnetic layer
the superconductivity is reentrant with
{\it temperature} $T$: as one lowers $T$ the
system turns superconducting, and when $T$ is further lowered it turns
normal again. This behavior is reflected in
the condensation free energy and the pair potential, which
vanish both above the upper transition and  below the lower one.
The transition is strictly reentrant: the low and high temperature
phases are the same. 
The entropy further reveals a range of temperatures where the superconducting state
is {\it less ordered} than the normal one. 

\end{abstract}


\maketitle
More than thirty years ago, reentrant superconductivity associated with 
magnetic ordering was first observed  in the ternary rare-earth 
compounds ErRh$_4$B$_4$ and 
HoMo$_6$S$_8$\cite{cite:fertig,cite:moncton,cite:ott,cite:crabtree,cite:lynn}.
On cooling, 
these materials first become superconducting at a critical 
temperature $T_{c2}$. Upon further cooling,  inhomogeneous magnetic
order sets in. This ordering coexists 
with superconductivity\cite{cite:buzdin4}
over a very narrow $T$ range. This onset is  nearly immediately\cite{range} 
followed 
by that of long-range ferromagnetic order, which
entails the destruction of superconductivity, at a second 
critical temperature $T_{c1}$. Thus,
the reason for
the disappearance of the superconductivity at $T_{c1}$ is
essentially the presence of the magnetism.
That 
nonuniform magnetic ordering can appear in the presence of superconductivity
is consistent with the prediction made by 
Anderson and Suhl\cite{cite:anderson}.
Reentrant superconductivity of a
different kind is also 
found in ferromagnet/superconductor (F/S) layered 
heterostructures\cite{cite:buzdin4}. 
On increasing the thickness, $d_F$, of the ferromagnet layers in
such structures, while keeping the thickness 
of the superconductor layers constant, the superconductivity 
may disappear  for a certain range of 
thickness ($d_{F1}<d_F<d_{F2}$) and then return for larger $d_F$ 
($d_F>d_{F2}$).

The purpose of this Letter is to show that  
superconductivity in  F/S nanostructures which is
reentrant {\it with temperature} can occur under some circumstances, 
when the magnetic
structure is non-uniform.  
That is, for certain types of ferromagnets, the Cooper pair amplitude 
in such structures 
can be nonvanishing in a range 
$T_{c1}<T<T_{c2}$, with $T_{c1}>0$. 
Specifically, we have found that this reentrance occurs
in F/S bilayers where the magnetic order
of the F layer  is of the spiral type, as in  Holmium\cite{sosnin}.
The reentrance we find is very different from that in ErRh$_4$B$_4$ or 
HoMo$_6$S$_8$. There, the high $T$ phase is paramagnetic and the low $T$ phase
is ferromagnetic. In our case, the magnetic
order remains unchanged: it is the same above $T_{c1}$, below $T_{c2}$,
and {\it in between}. Reentrance occurs also\cite{oned} in some
quasi one dimensional superconductors, but there 
the low $T$ phase is insulating.
In our case, we have {\it strict reentrance}: the lowest $T$
and highest $T$ phases are the same, while in the entire range
in between, superconductivity and magnetism harmoniously coexist. This 
is unusual. Superconducting reentrance is also found  in granular
films\cite{granular}: it is not due
to magnetism but it involves the turning on and off of the intergrain
Josephson coupling. Here, we are able to evaluate the thermodynamic
functions of the system as it undergoes the transitions,  
and from their behavior one can glimpse the reasons
for the occurrence of the reentrance. The balance between the 
internal energy of the system and its entropy can result in a 
situation 
where the entropy of the thermodynamically stable superconducting state 
is higher than that of the normal state.

Extensive 
theoretical\cite{cite:buzdin2,cite:khusainov,cite:fominov,cite:buzdin3,cite:buzdin4,hv1,hv2}
work indicates that
the origin of $d_F$ reentrant superconductivity in F/S 
nanostructures can be traced to the damped oscillatory nature of the 
Cooper pair wave functions in ferromagnets\cite{cite:demler,cite:khov1}.
Qualitatively, when a Cooper pair enters into an F region, it decays and the electron with 
magnetic moment parallel 
to the internal exchange field $\mathbf{h}$ lowers its energy by 
an amount proportional to $h$, while the other electron 
with opposite spin raises its energy by the same amount. 
Then, the kinetic energy of
each electron changes and as a result\cite{cite:demler}
the Cooper pair entering 
into an F region acquires a spatially dependent phase in the 
F layer. 
This propagating character of the Copper pair leads to interference 
between the transmitted pairing wave function through the F/S interface 
and the reflected wave from the opposite surface of the ferromagnet. 
Experimentally, the reentrant behavior of superconductivity 
with $d_F$ has been observed 
and confirmed in Nb/Cu$_{1-x}$Ni$_x$ bilayers and Fe/V/Fe 
trilayers\cite{cite:garifullin,cite:zdravkov1,cite:zdravkov2}.
However, in the work we present here, reentrance occurs
with temperature, rather than just
with geometry. Thus, although it is already known\cite{cite:linder}
that the nonuniform Ho structure has strong effects
in the S/F proximity phenomena,  
no 
$T$ reentrance results have been predicted or observed. 

In the rest of this paper, we will first review our methods 
as applied to Holmium/superconductor (Ho/S) structures
and then 
discuss the microscopic behavior of the pair
amplitudes as well as the thermodynamic 
quantities.    
The approach we use here is based on exact, self-consistent,
diagonalization of the Bogoliubov-de Gennes (BdG)\cite{cite:degennes}
equations for clean  
F/S structures.
This approach not only has the virtue of being very general 
but 
is also able to 
describe short wavelength oscillations, which is important for small structures. 
The self consistent methods we use to diagonalize the BdG equations have
been extensively described in the literature (see e.g. Ref.~\onlinecite{hv3}
and references therein) and  details will not be given here, except where
crucial.

\begin{figure}
\includegraphics[width=0.5\textwidth] {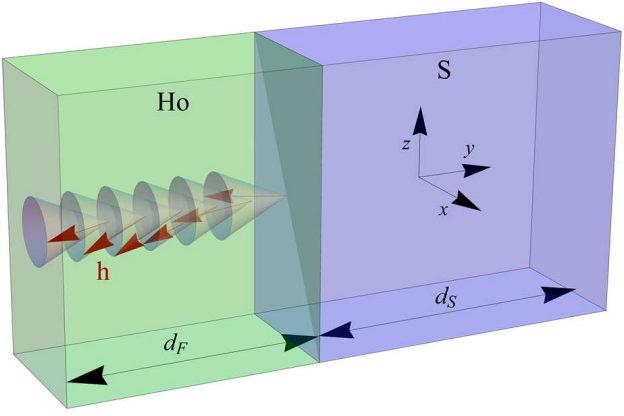} 
\caption{(Color online) Schematic of the ferromagnet (Ho) - superconductor 
(S) bilayer studied.  The conical ferromagnet has a 
spiral magnetic structure described by an exchange field ${\bf h}$, 
(see text). The system is infinite in the $x-z$ plane and $y$ is normal 
to the interfaces. 
} 
\label{figstruct} 
\end{figure}

The geometry of the Ho/S system we consider is depicted 
schematically in Fig.~\ref{figstruct}. The $y$ axis is normal to the
layers. 
The system is assumed to be infinite in the $x$-$z$ plane and has a total length 
$d$ in the $y$ direction.  
The S layer in our 
assumed Ho/S system is a conventional $s$-wave superconductor with 
thicknesses $d_S$ and a Ho layer of thickness $d_F$. As in previous
work, the magnetic structure is described via a local exchange field 
$\mathbf{h}$  which in this case is of the form: 
$\mathbf{h}=h_0\left\{\cos\theta\mathbf{\hat y}+\sin\theta\left[ \sin\left(\frac
{\varphi y}{a}\right)\mathbf{\hat x}+\cos\left(\frac{\varphi y}{a}\right)\mathbf
{\hat z}\right]\right\}$, where for Ho we have\cite{cite:linder,sosnin}
 $\theta=4\pi/9$ and $\varphi=\pi/6$.
We will take $a$, the lattice constant, as our unit of length and assume 
throughout that the system is below the temperature ($21$ K) at which, 
$\theta$ switches from $\pi/2$ to $4\pi/9$, i.e.  
Ho becomes ferromagnetic.

The effective Hamiltonian, $\mathcal{H}_{eff}$, that we use to model our 
Ho/S system 
takes the form
\begin{eqnarray}
\mathcal{H}_{eff} &=& \int
d^3r\Bigl\lbrace\displaystyle\sum\limits_{\alpha}\psi_{\alpha}^{\dagger}(\mathbf{r})\left(-\frac
{\boldsymbol\nabla^2}{2m^*}-E_f\right)\psi_{\alpha}(\mathbf{r})
 \nonumber \\
&&
+\frac{1}{2}\left[\displaystyle\sum\limits_{\alpha,\beta}(i\sigma_y)_{\alpha\beta}
\Delta(\mathbf{r})\psi_{\alpha}^{\dagger}(\mathbf{r})\psi_{\beta}^{\dagger}(\mathbf{r})+
h.c.\right]
 \nonumber \\
&& -\displaystyle\sum\limits_{\alpha,\beta}\psi_{\alpha}^{\dagger}(\mathbf{r})(\mathbf{h}
\cdot\boldsymbol\sigma)\psi_{\beta}(\mathbf{r})\Bigr\rbrace,
\end{eqnarray}
where $\Delta(\mathbf{r})$ is the usual singlet 
pair potential; $\psi_{\alpha}^{\dagger}$ and 
$\psi_{\alpha}$ are the creation and annihilation operators with spin 
$\alpha$ respectively; $E_f$ is the Fermi energy and $\boldsymbol{\sigma}$ are 
the Pauli matrices. 
To recast the $\mathcal{H}_{eff}$ into  diagonal form, we apply a  
generalized Bogoliubov transformation,
$\psi_{\alpha}(\mathbf{r})=\sum\limits_n\left[u_{n\alpha}(\mathbf{r})\gamma_n+
v_{n\alpha}^\ast(\mathbf{r})\gamma_n^{\dagger}\right]$,
where the quantum number $n$ enumerates the 
quasiparticle ($u_{n\alpha}$) and quasihole ($v_{n\alpha}$) spinors.
The $\gamma_n$ and $\gamma_n^\dagger$ are the Bogoliubov quasiparticle annihilation
and creation operators respectively.
By making use of  the commutation relations, 
$\left[\mathcal{H}_{eff},\gamma_n\right]=-\epsilon_n\gamma_n$ and   
$\left[\mathcal{H}_{eff},\gamma_n^{\dagger}\right]=\epsilon_n\gamma_n,$
one obtains the BdG equations in matrix form. In the geometry chosen,
the dependence of the wavefunctions on the $x$ and $z$ variables leads
to an obvious phase factor that can be canceled out. This results in
a set of quasi one dimensional problems of the form:
\begin{align}
\begin{pmatrix}H_e-h_z&-h_x+ih_y&0&\Delta \\ -h_x-ih_y&H_e+h_z&-\Delta&0 \\ 
0&-\Delta^{\ast}&-H_e+h_z&h_x+ih_y \\ \Delta^{\ast}&0&h_x-ih_y&-H_e-h_z\end{pmatrix} 
\nonumber \\
\times\begin{pmatrix}u_{n\uparrow} \\ u_{n\downarrow} \\ v_{n\uparrow} \\ 
v_{n\downarrow}\end{pmatrix}=\epsilon_n\begin{pmatrix}u_{n\uparrow} \\ 
u_{n\downarrow} \\ v_{n\uparrow} \\ v_{n\downarrow}\end{pmatrix},
\end{align} 
where $H_e\equiv-(1/2m^*)({\partial^2}/{\partial
y^2})+\epsilon_\perp-E_f$,
with $\epsilon_\perp$ being the kinetic energy associated with the
transverse direction.
Thus the spatial dependence of the amplitudes is only on $y$. The
exchange field ${\bf h}(y)$ in Ho  is nonvanishing only in the F region and precesses
as given above (see also Fig.~\ref{figstruct}). 
The pair potential  must be determined 
self-consistently by solving the BdG equations together with the
condition,
\begin{equation}  
\Delta(y)=\frac{g(y)}{2}{\sum_n}^\prime 
\left[u_{n\uparrow}(y)v_{n\downarrow}^\ast(y)-u_{n\downarrow}(y)v_{n\uparrow}^\ast(y)
\right]\tanh(\frac{\epsilon_n}{2T}),
\label{op}
\end{equation}
where 
$T$ is the temperature,  and $g(y)$ is 
the usual BCS coupling constant associated with a contact potential that exists only in the S region.  
The prime on the sum implies that only states corresponding to 
positive energies below the ``Debye'' cutoff  
$\omega_D$ are included. 
The self consistent diagonalization is achieved as in the previous
work mentioned above, the only difference being that the matrices to be
diagonalized are in this case unavoidably complex. 

From the self consistent results one can evaluate immediately the pair
amplitudes and, as explained below, the thermodynamic quantities. The
transition temperatures can be most conveniently evaluated by 
a linearization method\cite{cite:pbov}. Near the
transition temperature, the equation for $\Delta$ can be 
written as
$\Delta_i=\sum_qJ_{iq}\Delta_q$,
where $\Delta_i$ are the expansion coefficients with respect to the orthonormal 
basis, $\phi_i(y)=\sqrt{2/d}\sin(i\pi y/d)$, and $J_{iq}$ is given as 
$J_{iq}\equiv(J_{iq}^u+J_{iq}^v)/2$, where
\begin{equation}
J_{iq}^u=\gamma\int d\epsilon_{\perp}\displaystyle\sum\limits_n
\left[\tanh\left(\frac{\epsilon_n^{u,0}}{2T}\right)\displaystyle\sum\limits_m 
\frac{F_{qnm}^\ast F_{inm}}{\epsilon_n^{u,0}-\epsilon_m^{v,0}}\right] 
\end{equation}
\begin{equation}
J_{iq}^v=\gamma\int d\epsilon_{\perp}\displaystyle\sum\limits_n
\left[\tanh\left(\frac{\epsilon_n^{v,0}}{2T}\right)\displaystyle\sum\limits_m 
\frac{G_{qnm}G_{inm}^\ast}{\epsilon_n^{v,0}-\epsilon_m^{u,0}}\right]
\end{equation}
Here $\gamma=\gamma_0/4\pi D$ with $\gamma_0$ being the dimensionless coupling 
constant in S; $D$ is the total dimensionless thickness of the 
structure, $D \equiv k_{fS} d$, and $k_{fS}$ is the Fermi 
wavevector in S. We take $k_{fS}=1/a$; 
$\epsilon_n^{u(v),0}$ are unperturbed particle(hole) energies; 
and
$F_{inm}=\pi\sqrt{2d}\sum_{pq}\left(u_{np\uparrow}^0u_{mq\downarrow}^0-u_{np
\downarrow}^0u_{mq\uparrow}^0\right)K_{inm}$, 
$G_{inm}=\pi\sqrt{2d}\sum_{pq}\left(v_{np\uparrow}^0v_{mq\downarrow}^0-v_
{np\downarrow}^0v_{mq\uparrow}^0\right)K_{inm}$,
where $K_{inm}\equiv\int_0^ddyg(y)\phi_i(y)\phi_n(y)
\phi_m(y)$. 
The $u_{np}^0$ and $v_{mq}^0$ are the expansion coefficients of the 
unperturbed ($\Delta=0)$ 
particle (hole) amplitudes in terms of the basis set. 

This linearization method is easily used to evaluate the transition temperature.
As explained in Ref.~\onlinecite{cite:pbov}, one simply has to find the
largest eigenvalue, $\lambda$, of the matrix $J_{iq}$ and see if it is greater 
or smaller than unity: in each case one  is, respectively,
in the superconducting or the normal state. The transition temperatures
are those at which the largest eigenvalue changes from  greater to
smaller than unity: one finds $T_c$ by evaluating $\lambda$ as a function of $T$.  
In the usual case
$\lambda$ is smaller than unity when 
$T$ is larger than $T_c$. In a reentrant case with superconductivity in the 
range $T_{c1}<T<T_{c2}$, we find 
$T_{c1}$ by increasing $T$ from zero until $\lambda>1$ and  
$T_{c2}$ by $decreasing$ $T$ from above $T_{c2}$ until $\lambda>1$. 
\begin{figure}
\includegraphics[width=0.48\textwidth] {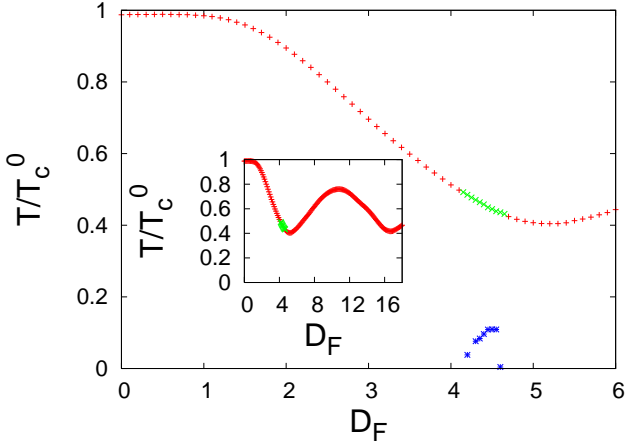} 
\caption{(Color online) Calculated transition temperature $T_c$, normalized by 
$T_c^0$ (see text), 
vs. the dimensionless ferromagnet width, 
$D_F (\equiv d_F k_{fS})$.  
Main plot: The 
upper points ((red) + , (green) $\times$ signs) are 
the usual critical temperature ($T_{c2}$),
leading  to the superconducting state  
as $T$ is lowered.
In the region $4 \lesssim D_F \lesssim 5$  (highlighted by the (green)
$\times$ signs) a second transition back to the normal 
state appears at the (blue) star points forming the 
lower ``dome". The inset shows a broader range of
magnet widths, revealing the overall periodicity of $T_{c2}$.} 
\label{figtc} 
\end{figure} 

In all results given here, the thickness of the S layer is fixed 
at  $d_S=(3/2) \xi_0$, where $\xi_0$ is the 
usual BCS coherence length in S. 
We take $\xi_0=100 k_{fS}^{-1}$, and vary $d_F$. 
The magnitude of 
$h$ is 
$0.15E_f$. 
Results for the transition
temperature, normalized to the bulk transition temperature $T_c^0$ of S, 
are shown in 
Fig.~\ref{figtc}, plotted as a function of $D_F\equiv d_F k_{fS}$. In the inset, we see 
that the
overall behavior of $T_c$ consists of the expected damped oscillations 
with approximately the $D_F$
periodicity of 
the spiral magnetic structure (twelve, in our units). The main
plot shows in more detail the structure near the first minimum. There
we see also  
a lower small dome-shape plot ((blue) stars) with a maximum 
at $D_F \approx 4.5$.
The system is in the normal phase inside the dome and, at constant
$D_F$, it is in the 
superconducting phase between the two curves.
In the $D_F$ range including the dome, the system, upon
cooling,  first becomes superconducting at a 
higher temperature $T_{c2}$, and with further cooling, returns to 
the normal phase at a lower temperature $T_{c1}$.

\begin{figure} 
\includegraphics[width=0.48\textwidth] {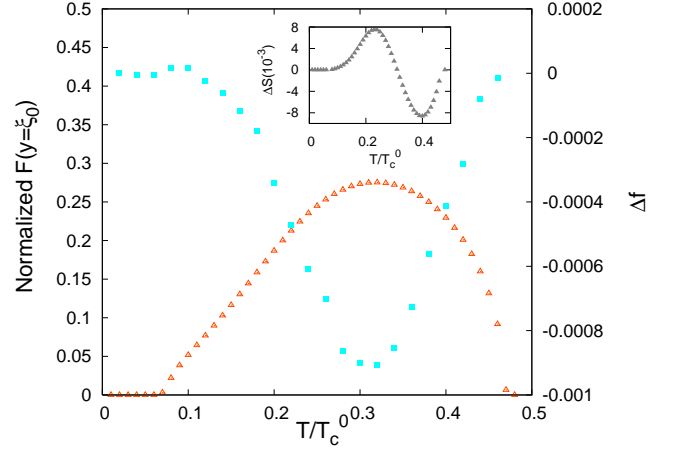}
\caption{ (Color online). Pair amplitude and 
thermodynamic functions. All quantities are plotted vs. $T/T_c^0$.
In the main plot, the (red) triangles and left vertical scale
display 
the normalized (see text) singlet Cooper pair amplitude $F(Y)$, 
one correlation length 
inside S. This quantity vanishes at the upper transition 
temperature 
($T_{c2}\approx
0.47 T_c^0$) and again at the lower transition $T_{c1}\approx 0.07 T_c^0$.  
The (blue) squares and right scale are
the normalized (see text) condensation free energy, $\Delta f$.
The vanishing of $\Delta f$  
at the upper and lower transitions
is clearly seen. The inset shows the normalized
entropy difference $\Delta S\equiv -({d\Delta f }/{d(T/T_c^0))}$. 
}
\label{figcombo}
\end{figure}

In Fig.~\ref{figcombo} we display additional 
direct evidence confirming the
existence of the reentrant behavior and showing its properties. All
results in the figure are for a system in the reentrant region, with
$D_F=4.3$,  and  are plotted vs. $T/T_c^0$ .
We consider first (main plot, (red) triangles, left vertical scale),   
the Cooper pair amplitude $F(y)$
defined by $\Delta(y) \equiv g(y) F(y)$ (see Eqn.~(\ref{op})).
The quantity shown is  
$F(y=\xi_0)$,  normalized to its bulk value in S,
at a position one coherence length inside S. 
This amplitude vanishes below $T_{c1}$ and above $T_{c2}$, 
with the values of $T_{c1}$ and $T_{c2}$  agreeing with those
previously found: 
we can see from Fig.~\ref{figtc}, $T_{c1}\approx 0.07 T_c^0$ and 
$T_{c2}\approx 0.47 T_c^0$ at 
$D_F=4.3$.
The continuity of the pair amplitude at $T_{c1}$ and
$T_{c2}$ 
also indicates that the transitions are of second order.

In the rest of Fig.~\ref{figcombo}   
the thermodynamics of the transitions, which follows from the
free energy, is shown.  Using a standard formalism \cite{cite:pbov,cite:kosztin},
we calculated 
$F_S$, the free energy of the whole system
in the self consistent state, and $F_N$, the normal state
($\Delta \equiv 0$) free energy. 
The normalized condensation free energy 
$\Delta f \equiv (F_S-F_N)/(2 E_0)$ ($E_0$ is the 
condensation energy of bulk S material at $T=0$) 
is then plotted in the main part of Fig.~\ref{figcombo} ((blue) squares,
right scale). 
Both $F_S$ and $F_N$ 
are monotonic and have negative
curvature with $T$ as required by thermodynamics, but their difference 
is nonmonotonic. Although 
$\Delta f$ is  small compared to its bulk value,
we can still identify the  two transition 
temperatures $T_{c1}$ and $T_{c2}$ 
from this plot.
Their values
are again in agreement, within 
numerical uncertainty, 
with those found from the pair amplitudes
and from direct calculation.
The system  is in the superconducting state when 
the $T$ falls in the range $T_{c1}<T<T_{c2}$. 
As $T_{c1}$ is approached from above or $T_{c2}$ from below, the solution
with $\Delta \nequiv 0$ disappears (as seen in the amplitude plot, (red) 
triangles),
and the two free energies coincide: this is just what happens in ordinary BCS
theory as the transition is approached from below.  
The minimum condensation free energy occurs at $T_{m}\approx 0.32 T_c^0$ 
which coincides with the location of the maximum pair amplitude.
We also evaluated the entropy  in the normal and superconducting states 
via textbook formulas. 
The normalized\cite{cite:pbov} entropy difference for the same case 
is shown in the inset of 
Fig.~\ref{figcombo}.
It confirms that the system indeed undergoes  second order phase transitions at 
both $T_{c1}$ and $T_{c2}$. 
Unlike in a bulk superconductor, or in non-reentrant
structures\cite{cite:pbov}, there is now a range of $T$  
($T_{c1}<T<T_{m}$) where the superconducting state is less 
ordered than the normal one, and the entropy helps maintain the 
superconductivity. 

What is the physics behind this $T$ reentrance?
For F/S bilayers with a uniform ferromagnet, the 
superconductivity disappears for a certain range of $d_F$. 
This disappearance is due to  the 
oscillating Cooper pair amplitude. 
Now, the spiral magnetization in Ho introduces an oscillating magnetic order.
Both the magnetic structure and the superconductivity are nonuniform,
consistent with the prediction in 
Ref.~\onlinecite{cite:anderson} 
that superconductivity may coexist with nonuniform 
magnetic order. 
Thermodynamically, we have here a subtle example of entropy-energy competition.
In the range $T_{m}<T<T_{c2}$, $\Delta f$ and $\Delta S$
behave qualitatively as they do for an ordinary\cite{tinkham} 
bulk superconductor
in the region $0<T<T_c$, (although they are  much smaller). 
In either case $\Delta S$  vanishes at both ends of the range and has a
minimum at a finite $T$ in between. But in our case 
$T_{m}$ is nonzero. For  $T<T_{m}$, $\Delta S$ turns positive because
of the oscillatory nature of the pair potential. 
The superconducting state becomes then the  higher entropy phase: 
the roles of the N and S phases are thus reversed, the pair
potential begins to decrease, and this leads inexorably to
the lower transition at $T_{c1}$, and to the reentrance into the same N phase.

We have already seen above the clear differences between this entropy
competition driven situation and other singlet superconducting 
$T$ reentrance cases associated with field induced
situations. Temperature reentrance involving long
range magnetic order has been long known to occur in spin glasses\cite{sg},
but the lowest $T$ and high $T$ phases (spin glass and paramagnetic respectively)
are not the same. Somewhat similar but even more 
complicated situations occur 
in liquid crystals and may be a general property of\cite{fru} frustrated
systems. But a survey would take us too far afield.



To our knowledge, this effect has not been searched for. The predicted
range of $T$ needed, down to about 0.1 $T_c^0$ should pose no difficulty.
The best course should be to fabricate samples of varying $d_F$, verify
the $T_c$ oscillations (see Fig.~\ref{figtc} inset) and then search for
reentrance for $d_F$ {\it near a minimum} of the  $T_c$ vs $d_F$ curve, where the
phenomenon is predicted to occur. (This is possibly because such minima
are associated with fragility of the superconducting state). 
It has proved experimentally feasible\cite{fsfexp} to study 
the $T$ induced $0-\pi$ state transitions in S/F/S trilayers,
which are related to a different effect\cite{cite:pbov} also
involving nontrivial pairing correlations. 
Thus, that difficulties in sample making are not 
insurmountable.

In conclusion, we predict
that F/S bilayers with an inhomogeneous conical magnetization will exhibit 
reentrant superconductivity with $T$, in addition to $d_F$. 
Thus,  superconductivity exists for 
$T_{c1}<T<T_{c2}$ with nonzero $T_{c1}$ under some conditions.
We have shown clear evidence for this by self consistently determining the
critical temperature-thickness phase diagram, and 
the $T$ dependence of the pair amplitude. The thermodynamics were
investigated via the free energy,
revealing a range of temperatures in which the normal state is lower in 
entropy than the superconducting state. 



\begin{thebibliography}{00}
\bibitem{cite:fertig} W. A. Fertig {\it et al}., Phys. Rev. Lett. {\bf 38}, 987 (1977).
\bibitem{cite:moncton} D. E. Moncton {\it et al}., Phys. Rev. Lett. {\bf 39}, 1164 (1978).
\bibitem{cite:ott} H. R. Ott {\it et al}., J. Low Temp. Phys. {\bf 33}, 1/2 (1978).
\bibitem{cite:crabtree} G. W. Crabtree {\it et al}., Phys. Rev. Lett. {\bf 49}, 1342 (1982).
\bibitem{cite:lynn} J. W. Lynn {\it et al}., Phys. Rev. B {\bf 31}, 5756 (1985).
\bibitem{cite:buzdin4} A. I. Buzdin, Rev. Mod. Phys. {\bf 77}, 935 (2005).
\bibitem{range} The $T$ range over which inhomogeneous
magnetism and superconductivity coexist in ErRh$_4$B$_4$ is only 0.1 K;
in HoMo$_6$S$_8$ even less: see p. 2 of Ref.~\cite{cite:buzdin4}.
\bibitem{cite:anderson} P. Anderson and H. Suhl, Phys. Rev. {\bf 116}, 898 (1959).
\bibitem{sosnin} I. Sosnin {\it et al.}, \prl {\bf 96}, 157002 (2006).
\bibitem{oned} See e.g. Brossard {\it et al} \prb {\bf 42}, 3935 (1990).
\bibitem{granular} This has been long known: see e.g. 
T.H. Lin {\it et al}, \prb {\bf 29}, 1493 (1984).
\bibitem{cite:buzdin2} Z. Radovic {\it et al}., Phys. Rev. B {\bf 44}, 759 (1991).
\bibitem{cite:khusainov} M. G. Khusainov and Y. N. Proshin, Phys. Rev. B {\bf 56}, R14283 (1997).
\bibitem{cite:fominov} Y. V. Fominov, N. M. Chtchelkatchev, and A. A. 
Golubov, Phys. Rev. B {\bf 66}, 014507 (2002)
\bibitem{cite:buzdin3} I. Baladie and A. Buzdin, Phys. Rev. B {\bf 67}, 014523 (2003).
\bibitem{hv1} K. Halterman and O.T. Valls, \prb  {\bf 70}, 104516 (2004).
\bibitem{hv2} K. Halterman and O.T. Valls, \prb  {\bf 72}, 060514(R) (2005).
\bibitem{cite:demler} E. A. Demler, G. B. Arnold, and M.R. Beasley, Phys. Rev. B {\bf 55}, 15174 (1997).
\bibitem{cite:khov1} K. Halterman and O. T. Valls, Phys. Rev. B {\bf 65}, 014509 (2001).
\bibitem{cite:garifullin} I. A. Garifullin {\it et al}., Phys. Rev. B {\bf 66}, 020505(R) (2002).
\bibitem{cite:zdravkov1} V. Zdravkov {\it et al}., Phys. Rev. Lett. {\bf 97}, 057004 (2006).
\bibitem{cite:zdravkov2} V. Zdravkov {\it et al}., Phys. Rev. B {\bf 82}, 054517 (2010).
\bibitem{cite:linder} J. Linder, T. Yokoyama, and A. Sudbo, Phys. Rev. 
B {\bf 79}, 054523 (2009).
\bibitem{cite:degennes} P. G. de Gennes, {\it Superconductivity of Metals and Alloys} 
Addison-Wesley, Reading, MA, 1989).
\bibitem{hv3} K. Halterman and O.T. Valls, \prb {\bf 80}, 104502 (2009).
\bibitem{cite:pbov} P. H. Barsic, O. T. Valls, and K. Halterman, Phys. Rev. B {\bf 75}, 104502 (2007)
\bibitem{cite:kosztin} I. Kosztin {\it et al}., Phys. Rev. B {\bf 58}, 9365
(1998).
\bibitem{tinkham} See Fig.~3.3 in M. Tinkham, {\it Introduction to
Superconductivity}, Dover, Mineola, NY (1996).
\bibitem{sg} See e.g. G. Aeppli {\it et al} \prb 28, 5160 (1983).
\bibitem{fru} C.K. Thomas and H.G. Katzgraber, \pre {\bf 84}, 040101(R), (2011).
\bibitem{fsfexp} V.V. Ryazanov {\it et al.}, \prl {\bf 86}, 2427 (2001); J.W.A.
Robinson {\it et al.}, \prb {\bf 76}, 094522 (2007). 





\end{thebibliography}
\end{document}